\documentclass{appolb}
\usepackage{graphicx}

\usepackage{amsfonts}
\usepackage{amsbsy}
\usepackage{amscd}
\usepackage{amsopn}
\usepackage{amstext}
\usepackage{amsxtra}
\usepackage{color}
\usepackage{epsfig}

\usepackage{graphicx}
\usepackage{amssymb}
\usepackage{epsfig}
\usepackage{amsmath}
\usepackage{dcolumn}
\usepackage{bm}

\newcommand{\be}{\begin{equation}}
\newcommand{\ee}[1]{\label{#1} \end{equation}}
\newcommand{\ba}{\begin{eqnarray}}
\newcommand{\ea}[1]{\label{#1} \end{eqnarray}}
\newcommand{\nl}{\nonumber \\}

\begin{document}
\title{Jet mass fluctuations and\\fragmentation functions%
\thanks{Presented at the \textit{XXIII Cracow EPIPHANY Conference, 9--12 January 2017, Cracow Poland}}%
}
\author{Karoly Urmossy
\address{Institute of physics, Jan Kochanowski University\\15 Swietokrzyska Street,
PL-25406 Kielce, Poland}
}

\maketitle
\begin{abstract}
The effect of jet mass fluctuations on the fragmentation process is examined in the framework of a statistical hadronisation model. In this model, the fragmentation scale $Q^2$ is taken to be the virtuality of the leading parton, and jet mass fluctuations are accounted for through this quantity. The scale evolution of the model is treated in the $\phi^3$ theory with leading-order splitting function and one-loop coupling.  
\end{abstract}
  
\section{Introduction}
\label{sec:intro}
Energetic quarks and gluons produced in high-energy collisions create highly collimated bunches (jets) of hadrons. As this hadronisation (fragmentation) process cannot be handled by perturbation theory, usually phenomenological models or empirical formulas are used for the description of the distributions (fragmentation functions - FFs) of certain types of hadrons in jets initiated by a certain type of quark or gluon (parton). When the square of the total fourmomentum of the jet (the mass $M_{jet}^2 = P_{jet}^2$) is much smaller then the energy of the jet, the ``leading'' parton can be regarded as on-shell, and hadron distributions can be calculated as the convolution of hard cross sections and FFs (factorisation theorem \cite{bib:Fact,bib:css}).

\begin{figure}
\begin{center}
\includegraphics[height=0.18\textheight]{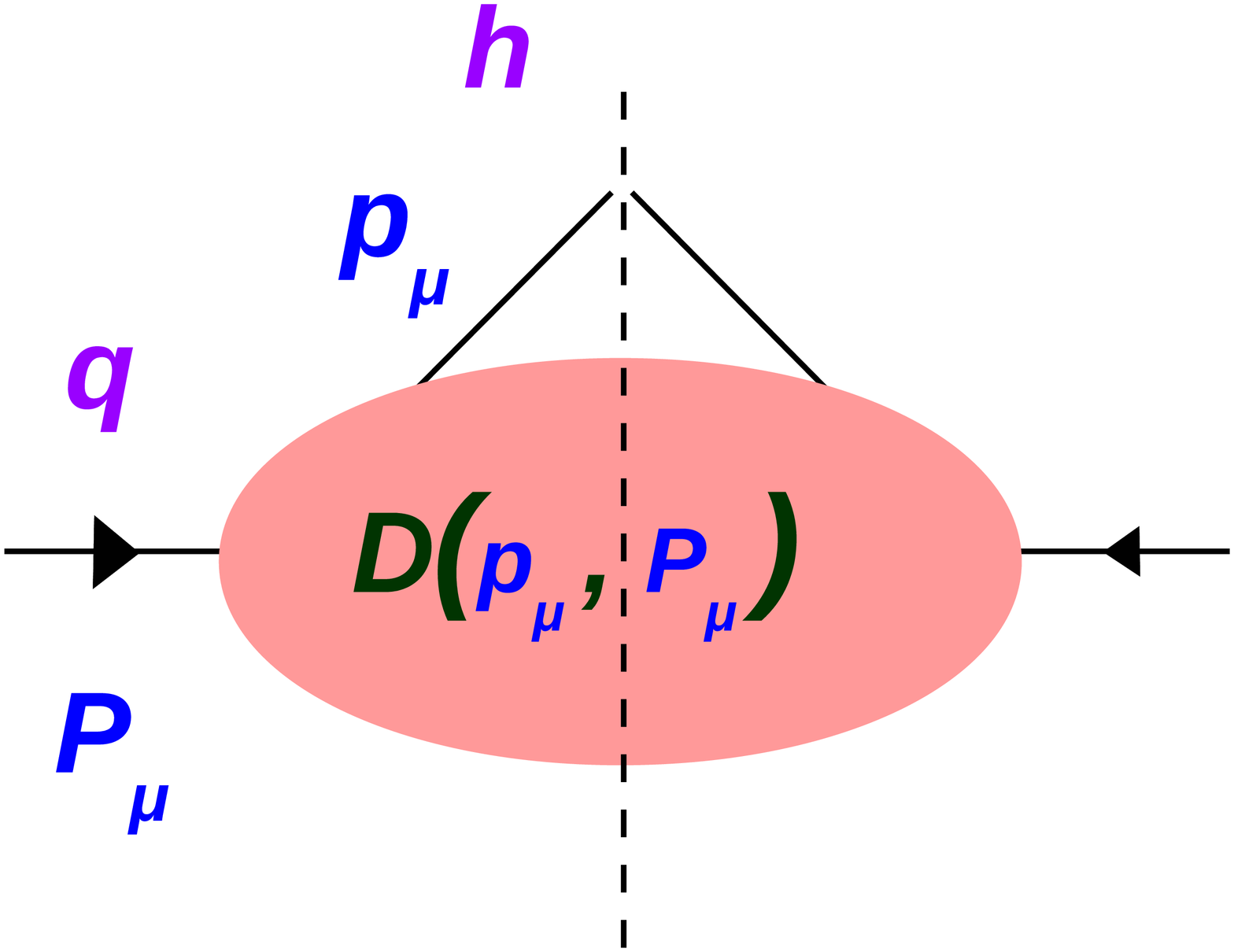}\hspace{10mm}
\includegraphics[height=0.18\textheight]{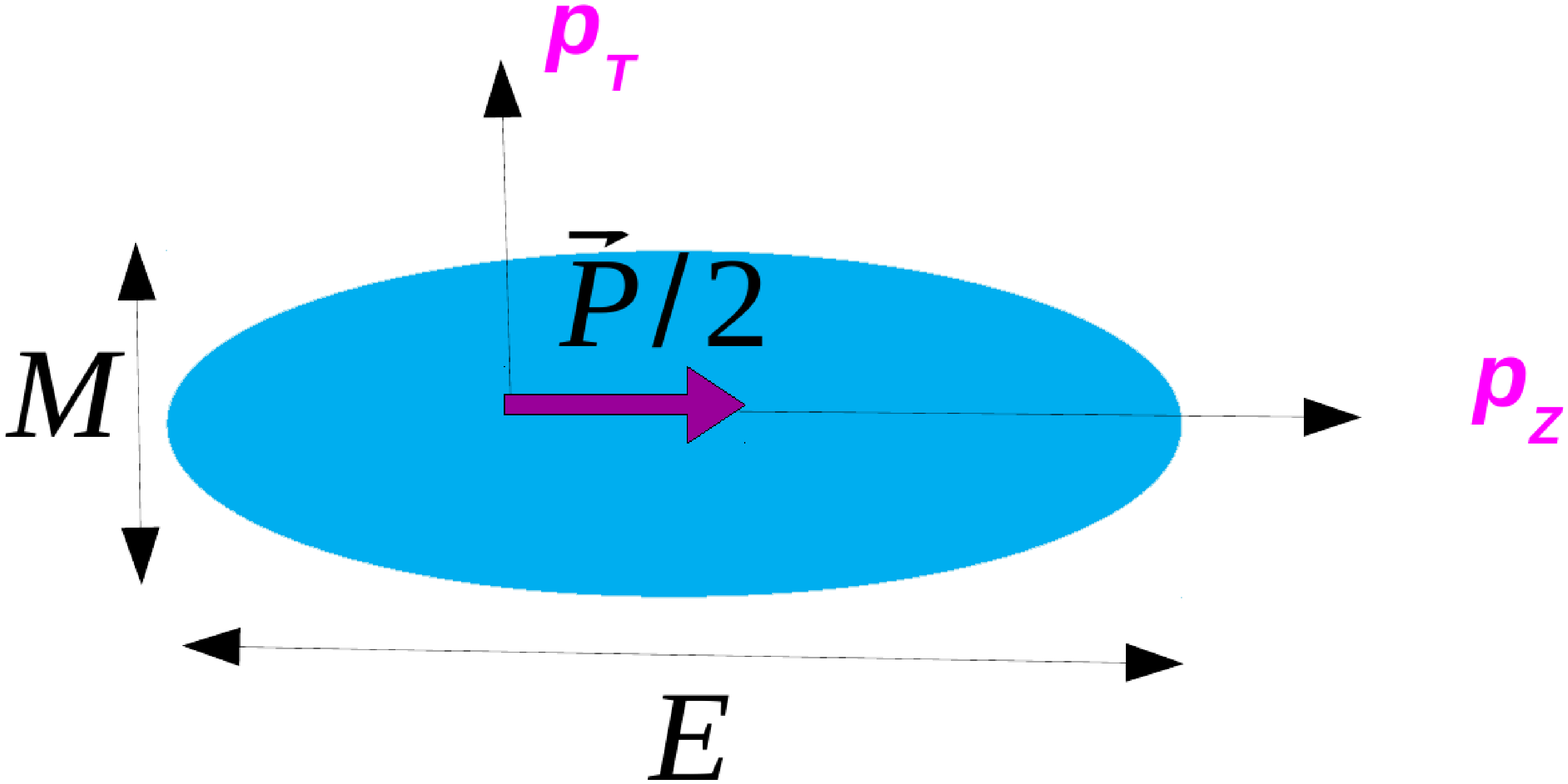} 
\end{center}
\caption{ \label{fig:FFgraph}
\textbf{Left,} subgraph of a jet with incoming initial parton $q$ of momentum $P_\mu$ and outgoing hadron $h$ of momentum $p_{\mu}$. \textbf{Right,} the phase space ellipsoid (with centre $\mathbf{P}/2$, longer axis $2a = E$ and smaller axis $2b = M$), available for hadrons in a jet of momentum $P_{\mu} = (E,\mathbf{P})$. In the limit of $|\mathbf{P}|\rightarrow E$, the ellipsoid shrinks, and Eq.~(\ref{eq2}) becomes a one-dimensional distribution of $\tilde x = p^0/E=x$.}
\end{figure}

However, there are cases when $M_{jet}$ is not negligable compared to $P^0_{jet}$. For example \cite{bib:UKpp3D}, the masses of jets fluctuate according to 
\be
\rho(M) \;\sim\;  \frac{\ln^b(M/\mu_0)}{M^c} \;
\ee{eq1}
in proton-proton (pp) collisions at $\sqrt s$ = 7 TeV \cite{bib:atlasM} and jet transverse momenta $P^{jet}_T \in$ [200--600] GeV/c. The masses of such jets are typically of the order of 60--100 GeV$/c^2$. According to \cite{bib:UKep3D,bib:UKpp3D}, in case of such ``fat'' jets, it is reasonable to parametrise FFs by the variable $\tilde x = 2P^{jet}_\mu p_h^\mu/M_{jet}^2$ ($p_h^\mu$ being the fourmomentum of the hadon)  and use $\tilde Q = M_{jet}$ as fragmentation scale instead of $x = p_h^0/P_{jet}^0$ and $Q = \theta_c P_{jet}^0$ ($\theta_c$ being the jet opening angle) which are most often used in the literature (eg. \cite{bib:MLLA1,bib:MLLA2,bib:MLLA3,bib:MLLA4,bib:MLLA5,bib:MLLA6,bib:dEnterria1,bib:dEnterria2}). On the one hand, looking at the schematic picture (Fig.~\ref{fig:FFgraph} left) of a jet, we only have two fourmomenta $P^{jet}_\mu$ and $p^h_\mu$ (in the spin-averaged case) to construct scalars from. As most created hadrons are pions, we may neglect the mass of the hadron $p_h^2 = m^2_\pi \approx 0$, and we are left with $P_{jet}^2 = M_{jet}^2$ and $P_{jet}^\mu p^h_\mu$. Thus, it is reasonable to use $\tilde x$ as the dimensionless variable, and $M_{jet}$ as the fragmentation scale. On the other hand, the width of the pha\-se\-spa\-ce  of hadrons inside the jet (Fig.~\ref{fig:FFgraph} right), allowed by energy-momentum conservation is equal to $M_{jet}$, rather than $\theta_c P_{jet}^0$.

If, however, we use $M_{jet}$ as fragmentation scale, we need to take into account its fluctuations when fitting experimental data. In Sec.~\ref{sec:relM}, I briefly summarise the statistical fragmentation model and its scale evolution discussed in detailes in \cite{bib:UKep3D,bib:UKpp3D}. Furtheremore, I obtaine the jet mass-averaged FF, and compare it experimental data measured in pp collisions at $\sqrt{s}$ = 7 TeV.  

\section{A statistical fragmentation model}
\label{sec:relM}
Although the scale evolution of FFs can be obtained using perturbation theory, a non-perturbative input is needed, namely the form of FFs at an initial scale $Q_0$. I use the statistical fragmentation model \cite{bib:UKep3D,bib:UKpp3D} for this purpose. In this model, the microcanonical ensemble is used (as in many cases in the literature \cite{bib:UKpp3D,bib:UKeeFF,bib:UKppFF,bib:UKeeFFstrange,bib:Becattini10,bib:Liu1,bib:Wibig3,bib:FHLiu,bib:Begun3,bib:Begun2,bib:Begun}) to account for the finiteness of the energy and multiplicity of jets. Besides, negative-binomial hadron multiplicity distributions are also taken into account, thus the single particle distribution inside jets of fix fourmomentum $P^\mu_{jet}$ becomes
\be
D\left(x,Q^2_0\right) \;=\; A_0 \left[1 + \frac{q_0-1}{\tau_0}\, x \right]^{-1/(q_0-1)} - B_0 \;,
\ee{eq2}
with $q_0$ and $\tau_0$ being parameters, $A_0$ is fixed by the normalisation condition, and $B_0 = A_0 \left[1 + \frac{q_0-1}{\tau_0} \right]^{-1/(q_0-1)}$ to ensure that $D\left( x=1,Q^2_0\right) = 0$ at the boundary of the phasespace ($x=1$).

For simplicity, I treat the scale evolution in the $\phi^3$ theory, where the DGLAP equation reads 
\be
\partial_t D(x,t) \;=\; g^2 \int\limits_x^1 \frac{dz}{z} D\left(\frac{x}{z},t\right) \Pi(z)\;
\ee{eq3}
with leading-order (LO) splitting function
\be
\Pi(x) \;=\; x(1-x) - \frac{1}{12}\delta(1-x)\;,
\ee{eq4}
scale variable $t = \ln\left(Q^2/\Lambda^2\right)$ and $\Lambda$ being the scale where $g^2(t) = 1/(\beta_0 t)$, the 1-loop coupling of the $\phi^3$ theory diverges; and $\beta_0$ being the first coefficient of the beta function of the $\phi^3$ theory. The solution is  
\be
D(x,t) \;=\; 
\int\limits_x^1 \frac{dz}{z} g\left(z,t\right) D\left(\frac{x}{z},t_0\right)
\ee{eq5}
with the kernel
\ba
g(x,t) &\sim& \delta(x-1) + \sum\limits_{k=1}^\infty \frac{b^k(t)}{k!(k-1)!} \sum\limits_{j=0}^{k-1} \frac{(k-1+j)!}{j!(k-1-j)!} \;\times \nl
&&\hspace{20mm} \times\;\; x \ln^{k-1-j}\left[\frac{1}{x}\right] \Big[ (-1)^j + (-1)^k x \Big]
\ea{eq6}
and I use the statistical fragmentation function given in Eq.~(\ref{eq2}) to serve as initial function at starting scale $t_0 = \ln\left(Q^2_0/\Lambda^2\right)$. 

\begin{figure}
\begin{center}
\includegraphics[width=0.9\textwidth]{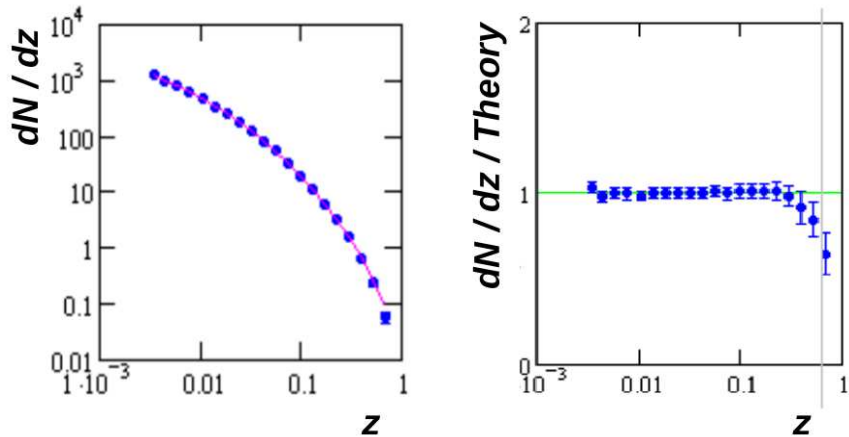} 
\end{center}
\caption{\textbf{Left,} comparison of the jet mass-averaged FF Eq.~(\ref{eq7}) and measured FF inside jets of $P^{jet}_T \in [400, 500]$ GeV/c in pp collisions at $\sqrt s$ = 7 TeV. The parameters of the initial fragmentation function Eq.~(\ref{eq2}) at scale $Q_0 = 1$ GeV are $q_0 = 1.275, \tau_0 = 0.02, \beta_0 = 0.1, \Lambda = 0.2$. Parameters of the mass distribution Eq.~(\ref{eq1}) are $b$ = 70, $c$ = 18 and $\mu_0$ = 1.4 GeV. \textbf{Right,} data over theory plot.
\label{fig:MassAver}}
\end{figure}

In order to compare this result to experimental data, we need to calculate the jet-mass-averaged FF
\be
\frac{dN}{dz} \;=\; 
\int dM \rho(M) D\left[z,\ln\left(\frac{M^2}{\Lambda^2}\right)\right]\;,
\ee{eq7}
as in case of available experimental datasets, the jet mass is not fixed. There is only one published dataset pair, in case of which, the kinematical properties of jets used when making the mass distribution and the fragmentation function coincides. This is the case of jets with transverse momenta $P^{jet}_T \in [400, 500]$ GeV/c in pp collisions at $\sqrt s$ = 7 TeV \cite{bib:atlasM,bib:atlasFFpp7TeV}. As Fig.~\ref{fig:MassAver} shows, in case of this dataset, smooth description of the measured FF can be achieved with Eq.~(\ref{eq7}). The parameters of the mass distribution were obtained in \cite{bib:UKpp3D} by fitting Eq.~(\ref{eq1}) to data in \cite{bib:atlasM}.

This result nicely supports the idea of using the jet mass as the fragmentation scale. This way, however, it would be advantageous to have experimental data on fragmentation functions inside jets of fixed mass instead of fixed energy or transverse momentum.

\subsection{Acknowledgement}
This work was supported by the Polish National Science Center grant 2015/19/B/ST2/00937.

\bibliographystyle{h-physrev3}
\bibliography{Urmossy_Epiphany2017}


\end{document}